\pdfoutput=1

\documentclass[11pt,twoside,a4paper,cmspaper,final,collab]{cms-tdr}
\def\svnVersion{379440}\def\cmsCernNoTag{CERN-EP/2016-204}\def\cmsCernDate{\today}\def\cmsMessage{Published in Physical Review D as \href{http://dx.doi.org/10.1103/PhysRevD.94.112004}{\doi{10.1103/PhysRevD.94.112004.}}}

\begin{document}\cmsNoteHeader{EXO-15-010}

\hyphenation{had-ron-i-za-tion}
\hyphenation{cal-or-i-me-ter}
\hyphenation{de-vices}
\RCS$Revision: 379431 $
\RCS$HeadURL: svn+ssh://svn.cern.ch/reps/tdr2/papers/EXO-15-010/trunk/EXO-15-010.tex $
\RCS$Id: EXO-15-010.tex 379431 2017-01-04 16:24:26Z kazana $
\newlength\cmsFigWidth
\ifthenelse{\boolean{cms@external}}{\setlength\cmsFigWidth{0.85\columnwidth}}{\setlength\cmsFigWidth{0.4\textwidth}}
\ifthenelse{\boolean{cms@external}}{\providecommand{\cmsLeft}{top\xspace}}{\providecommand{\cmsLeft}{left\xspace}}
\ifthenelse{\boolean{cms@external}}{\providecommand{\cmsRight}{bottom\xspace}}{\providecommand{\cmsRight}{right\xspace}}
\ifthenelse{\boolean{cms@external}}{\providecommand{\NA}{\ensuremath{\cdots}\xspace}}{\providecommand{\NA}{\ensuremath{\text{---}}\xspace}}
\ifthenelse{\boolean{cms@external}}{\providecommand{\CL}{C.L.\xspace}}{\providecommand{\CL}{CL\xspace}}
\ifthenelse{\boolean{cms@external}}{\providecommand{\cmsTableResize[1]}{\relax{#1}}}{\providecommand{\cmsTableResize[1]}{\resizebox{\columnwidth}{!}{#1}}}

\newcommand{\stau}{\ensuremath{\widetilde \tau _1}\xspace}
\newcommand{\ias}{\ensuremath{I_{\mathrm{as}}}\xspace}
\newcommand{\ih}{\ensuremath{I_{\mathrm{h}}}\xspace}
\newcommand{\dedx}{\ensuremath{\rd E/\rd x}\xspace}
\newcommand{\tof}{TOF}
\newcommand{\invbeta}{\ensuremath{1/\beta}\xspace}
\newcommand{\tkonly}{\textit{tracker-only}\xspace}
\newcommand{\tktof}{\textit{tracker+\tof}\xspace}
\newcommand{\multicharge}{\textit{multiply charged}\xspace}
\newcolumntype{x}{D{,}{\,\pm\,}{5,4}}
\cmsNoteHeader{EXO-15-010}
\title{Search for long-lived charged particles in proton-proton collisions at \texorpdfstring{$\sqrt{s}=13$\TeV}{sqrt(s)=13 TeV}}

\date{\today}

\abstract{
Results are presented of a search for heavy stable charged particles produced in proton-proton collisions at $\sqrt{s} = 13$\TeV
using a  data sample corresponding to an integrated luminosity of 2.5\fbinv
collected in 2015 with the CMS detector at the CERN LHC.
The search is conducted using signatures of anomalously high energy deposits  in the silicon tracker
and long time of flight measurements by the muon system.
The data are consistent with the expected background, and upper limits are set on the cross sections for production
of long-lived gluinos, top squarks, tau sleptons, and leptonlike long-lived fermions.
These upper limits are equivalently expressed as lower limits on the masses of new states;
the limits for gluinos, ranging up to 1610\GeV, are the most stringent to date.
Limits on the  cross sections for direct pair production of long-lived tau sleptons  are also determined.}

\hypersetup{%
pdfauthor={CMS Collaboration},%
pdftitle={Search for long-lived charged particles in proton-proton collisions at sqrt(s) = 13 TeV},%
pdfsubject={CMS},%
pdfkeywords={CMS, physics, long-lived particles}
}
\maketitle
\section{Introduction}

Many extensions of the standard model (SM) include
heavy long-lived charged particles that
might have high momentum, but speed
significantly smaller than the speed of light~\cite{Drees:1990yw, Fairbairn:2006gg, Bauer:2009cc}
and/or charge, Q, not equal to the elementary charge
$\pm 1e$~\cite{Kusenko:1997si, Koch:2007um,Schwinger:1966nj, Fargion:2005ep}.
Those particles with lifetimes greater than a few nanoseconds can
travel distances larger than the size of a typical collider detector and
appear quasi-stable like the pion or kaon.
These particles are generally referred to as heavy stable
charged particles (HSCPs) and can be singly ($\abs{Q}=1e$),
fractionally ($\abs{Q}<1e$), or multiply ($\abs{Q}>1e$) charged.
Without dedicated searches, HSCPs may be misidentified or unobserved,
since charged particle identification algorithms at hadron collider experiments
generally assume that particles
have speeds close to the speed of light and charges of ${\pm}1e$.
Additionally, HSCPs may be charged during only a part of their passage through detectors~\cite{Kraan:2004tz}
further limiting the ability of standard algorithms to identify them.

For HSCP masses greater than about 100\GeV,
a significant fraction of particles produced at the
LHC will have a relative velocity
$\beta \equiv v/c< 0.9.$  It is possible to distinguish $\abs{Q} \ge 1e$ particles
with $\beta < 0.9$
from light SM particles traveling close to the speed of light through
their higher rate of energy loss via ionization (\dedx) or
through their longer time of flight (\tof) to the outer detectors.
This paper describes a search for HSCPs using
the CMS detector in two ways:
(i)~requiring tracks to be reconstructed only in the silicon detectors, the \tkonly analysis;
(ii)~requiring tracks to be reconstructed in both the silicon detectors
and the muon system, referred to as the \tktof analysis.

\begin{figure}[bht]
 \centering
  \includegraphics[width=0.48\textwidth]{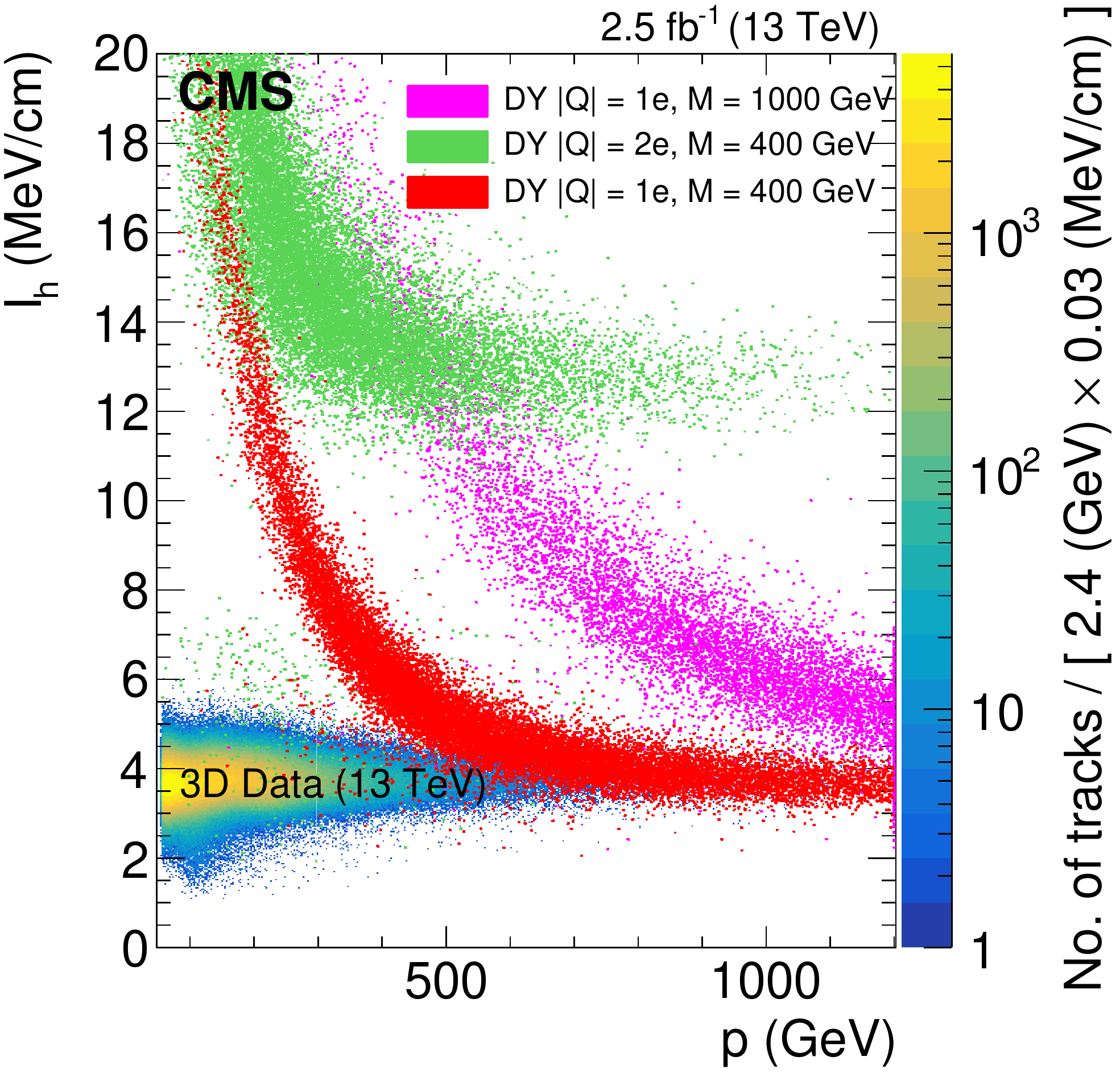}
 \caption{Distribution of the \dedx estimator, \ih (see Section~\ref{sec:dEdx}),
versus particle momentum for tracks in the 13\TeV data, and for simulation of HSCP for singly or multiply charged particles
with masses of 400 and 1000\GeV. The vertical scale shows the density of entries for data only.
   \label{fig:dedxdists}}
\end{figure}

The dependence of \dedx on the particle momentum is described by the Bethe-Bloch
formula~\cite{Agashe:2014kda}.
This dependence can be seen in Fig.~\ref{fig:dedxdists},
which shows the \dedx estimator versus momentum for
good quality (Section~\ref{sec:presel}) high transverse momentum ($\pt > 55$\GeV) tracks from data and
the generated Monte Carlo (MC) samples for  HSCP signals  with various charges.
In the momentum range of interest at the LHC (10--1000\GeV), SM particles
have nearly uniform ionization energy loss ($\approx$3\unit{MeV/cm}).
Searching for candidates with larger \dedx gives sensitivity
to massive particles with $\abs{Q} \geq 1e$.

Previous collider searches for HSCPs have been performed at
LEP~\cite{Barate:1997dr, Abreu:2000tn,Achard:2001qw,Abbiendi:2003yd},
HERA~\cite{Aktas:2004pq},
the Tevatron~\cite{Aaltonen:2009kea, Abazov:2008qu,Abazov:2011pf,Abazov:2012ina}
and the CERN LHC during Run~1 (proton-proton collisions with $\sqrt{s}$ up to 8\TeV)~\cite{Aad:2011hz, Aad:2011yf, Aad:2011mb,
Aad:2012pra, Aad:2013pqd,
Khachatryan:2011ts, Chatrchyan::2012dr, Chatrchyan:2012sp, Chatrchyan:2013oca}.
The results from these searches have placed significant bounds on
theories beyond the SM~\cite{Berger:2008cq,CahillRowley:2012kx}, such as lower limits at  95\% confidence
level (\CL) on the mass of long-lived gluinos (1300\GeV), top squarks (900\GeV),
and directly pair-produced tau sleptons (330\GeV).

In the present paper, results of searches for singly
and multiply charged HSCPs in 2.5\fbinv of data collected with
the CMS detector at $\sqrt{s} = 13$\TeV in 2015
are presented.
Similar limits on HSCPs were recently obtained by the ATLAS experiment~\cite{Aaboud:2016dgf,Aaboud:2016uth} using 3.2\fbinv of 13\TeV data collected in 2015.

\section{Signal benchmarks \label{sec:signals}}

The analyses described in this paper employ several
HSCP models as benchmarks, to account for a range
of signatures that are experimentally accessible.

The first type of signal consists of HSCPs that interact via the strong force
and hadronize with SM quarks to form $R$-hadrons~\cite{Fairbairn:2006gg, Bauer:2009cc}.
As in Ref.~\cite{Chatrchyan:2013oca}, events involving direct pair production of gluinos (\PSg) and
top squarks (\stone), with mass values in the range 300-2600\GeV, are generated according to the Split Supersymmetry
(Split SUSY) scenarios~\cite{Giudice:2004tc,ArkaniHamed:2004yi,Hewett:2004nw,Kilian:2004uj}.
Gluinos are generated assuming the squark mass is 10\TeV~\cite{ArkaniHamed:2004fb,Giudice:2004tc}.
In the region of parameter space where squarks are too heavy to be produced at the LHC,
the gluino-gluino production cross section and kinematic distributions depend only on
the gluino mass, thus the cross section limits are model-independent.
\PYTHIA~{8.153}~\cite{Sjostrand:2007gs}, with the underlying event tune CUETP8M1~\cite{Khachatryan:2015pea},
is used to generate the 13\TeV MC samples.
The fraction, $f$, of produced \PSg\  hadronizing into a \PSg-gluon
state is an arbitrary value of the hadronization model.
It determines the fraction of $R$-hadrons that are neutral at production.
For this search,
results are obtained for two different values of $f$, 0.1 and 0.5.
As in Ref.~\cite{Chatrchyan:2013oca}, two scenarios of $R$-hadron strong interactions with nuclear matter are considered.
The first scenario
follows the cloud model in Refs.~\cite{Kraan:2004tz, Mackeprang:2006gx},
which assumes that the $R$-hadron is surrounded by a cloud of colored, light constituents
that interact during scattering. Therefore, the $R$-hadron interacting inside the detector may change its charge sign.
The second scenario adopts a model of complete charge suppression \cite{Mackeprang:2009ad}
where the $R$-hadron becomes a neutral particle before it enters the muon system.
Both the \tkonly and \tktof analyses are used to search for these
signals, although only the \tkonly analysis is expected to have sensitivity
in the charge-suppressed scenario.
In the case of a discovery, a comparison of the numbers of events found
in the two analyses could give a hint about the nature of the new long-lived particle.

The second type of signal consists of HSCPs that behave like leptons.
The  minimal gauge mediated
supersymmetry breaking
(mGMSB) model~\cite{Giudice:1998bp} is selected as a benchmark for
leptonlike HSCPs. Production of quasi-stable sleptons
at the LHC can proceed either directly or via production of heavier
supersymmetric particles (mainly squarks and gluinos) that
decay and lead to
two sleptons
at the end of the decay
chain.
This latter process is dominant because
the direct production process is electroweak.
Direct production is relevant only
if squarks and gluinos are too heavy to be produced at the LHC.
The mGMSB model is explored using the SPS7 slope~\cite{Allanach:2002nj}, which
has the tau slepton (stau \stau) as the next-to-lightest
supersymmetric particle (NLSP).
The particle mass spectrum and the decay table are produced with the program
\textsc{isasugra}~7.69~\cite{Paige:2003mg}.
The  mGMSB model is characterized by six fundamental parameters.
The mGMSB parameter $\Lambda$, which corresponds to the effective supersymmetry breaking scale,
is varied from 31 to 510\TeV.
It is proportional to the sparticle masses.
The range of its values gives a tau slepton mass of 100 to 1600\GeV.
Other parameters are fixed to the following values.
The number of the messenger SU(5) multiplets $N_{\text{mes}}=3$ and their mass scale $M$ is set as
$M_{\text{mes}}/\Lambda=2$.
The ratio of the vacuum expectation values of the Higgs doublets is $\tan\beta=10$
and a positive sign of the higgsino mass term, $\mu > 0$, is assumed.
The large value of  the scale factor of the gravitino coupling, $C_{\text{grav}}=10000$, results in a long-lived \stau.
The SUSY mass spectrum produced is input to \PYTHIA\,{6.4}~\cite{Sjostrand:2007gs}
with the Z2*
tune~\cite{Chatrchyan:2013gfi} as the generator for a MC simulation at 13\TeV.
Two tau slepton samples are generated for each SUSY point: one with all processes
(labeled ``GMSB stau") and one with only direct pair production
(labeled ``Pair prod. stau").
The pair-produced stau includes only \stau, which is predominantly
$\widetilde \tau _R$ for these model parameters.
The direct production of long-lived stau is model independent. Both cross section and
kinematics depend only on the stau mass and the scan over the stau mass parameter
shows the effect of variations in center-of-mass energy and integrated luminosity.
The \tkonly and \tktof analyses are both used to search for these signals.

The last type of signal is based on modified Drell--Yan (DY) production of long-lived leptonlike fermions.
In this scenario, new massive spin-1/2
particles have arbitrary electric charge but are neutral under SU(3)$_{\text{Colour}}$
and SU(2)$_{\text{Left}}$, and therefore
couple only to the photon and the Z boson.
\PYTHIA v6.4~\cite{Sjostrand:2007gs} with the
Z2$^{\ast}$  tune~\cite{Chatrchyan:2013gfi} is used to generate these 13\TeV MC signal samples.
Simulations of events with leptonlike fermions
are generated with masses ranging from 100 to 2600\GeV
and for electric charges $\abs{Q}=1e$ and $2e$.

Different \PYTHIA tunes  were studied and
the effects on the kinematic distribution were negligible for the HSCPs considered.
The
\tkonly and \tktof analyses are both expected
to have sensitivity to $\abs{Q}=2e$ HSCPs.

In all signal samples, simulated minimum bias events are
overlaid with the primary collision to produce the effect of additional interactions in the same LHC beam
crossing (pileup).

\section{The CMS detector}

The central feature of the CMS~\cite{Chatrchyan:2008zzk} apparatus is a 3.8\unit{T} superconducting
solenoid of 6\unit{m} internal diameter.
Within the solenoid volume are
a silicon tracker, a lead tungstate crystal
electromagnetic calorimeter, and a brass and scintillator
hadron calorimeter, each composed of a~barrel and two
endcap sections. Outside the solenoid, forward calorimeters extend the
pseudorapidity ($\eta$) coverage provided by
the barrel and endcap detectors. Muons are measured in
gas-ionization detectors embedded in the steel flux-return
yoke outside of the solenoid.
The missing transverse momentum vector \ptvecmiss is defined
as the projection on the plane perpendicular to the beam axis of
the negative vector sum of the momenta of all reconstructed
particles in an event. Its magnitude is referred to as \ETmiss.

The silicon tracker, consisting of 1440
silicon pixel and 15\,148 silicon strip detector modules , measures charged particles
within the pseudorapidity range $\abs{\eta}< 2.5$.
Isolated particles of
transverse momentum
$\pt=100$\GeV and with
$\abs{\eta} < 1.4$ have track resolutions of 2.8\% in \pt
and 10\,(30)\mum\ in the transverse (longitudinal) impact
parameter \cite{TRK-11-001}.
Muons are measured in the pseudorapidity range $\abs{\eta}< 2.4$,
using three technologies: drift tubes (DTs),
cathode strip chambers (CSCs), and resistive-plate chambers (RPCs). Matching
muons to tracks measured in the silicon tracker results in
a relative transverse momentum resolution for muons with
$20 <\pt < 100\GeV$ of 1.3--2.0\% in the barrel and better
than 6\% in the endcaps. The \pt resolution in the barrel is
better than 10\% for muons with \pt up to 1\TeV~\cite{MUO-10-004}.

The first level (L1) of the CMS trigger system, composed of
custom hardware processors, uses information from the calorimeters
and muon detectors to select events of interest within
a fixed time interval of less than 4\mus. The high-level
trigger (HLT) processor farm further decreases the event rate
from around 100\unit{kHz} to less than 1\unit{kHz}, before data storage.
A more detailed description of the CMS detector, together with
a definition of the coordinate system used and the relevant
kinematic variables, can be found in Ref.~\cite{Chatrchyan:2008zzk}.

\subsection{\texorpdfstring{\dedx}{dE/dx} measurements \label{sec:dEdx}}

For the reconstructed track, information about \dedx
can be gained from measurements of ionization deposited
in layers of the pixel and silicon tracker.  The ionization charge
measured is compared with that expected from a Minimum-Ionizing Particle (MIP),
and its level of compatibility can provide a probability,
using a \dedx discriminator.
As in Ref.~\cite{Khachatryan:2011ts}, to distinguish SM particles from HSCP candidates
the \ias discriminator is used and is given by
\begin{equation}
 I_{as} = \frac{3}{N} \, \left(
   \frac{1}{12N} + \sum_{i=1}^N
   \left[
   P_i \, \left( P_i - \frac{2i-1}{2N} \right)^2 \right] \right),
\end{equation}
where $N$ is the number of measurements in the silicon-tracker detectors,
$P_i$ is the probability for a MIP to
produce a charge smaller or equal to that of the $i$th measurement
for the observed path length in the detector, and the sum is over the track
measurements ordered in terms of increasing $P_i$.

In addition, the \dedx of a track is estimated using a harmonic-2 estimator:
\begin{equation}
 I_{\mathrm{h}}= \biggl( \frac{1}{N_{85\%}} \sum_i^{N_{85\%}} c_{i}^{-2} \biggr)^{-1/2},
 \label{eq:HarmonicEstimator}
\end{equation}
where $c_{i}$ is the charge per unit path length in the sensitive part
of the silicon detector of the $i$th track measurement.  The harmonic-2 estimator
has units \MeV/cm
and the summation includes just the top 85\% of the charge measurements.
Ignoring the low charge measurements increases the resilience
of the estimator against instrumental biases.
This procedure is not necessary for \ias which is, by construction, robust against that type of bias.

The mass of a candidate particle
can be calculated~\cite{Chatrchyan:2013oca} from its momentum and its \ih \dedx estimate, based on the relation:
\begin{equation}
 I_{\mathrm{h}}= K\frac{m^2}{p^2}+C,
 \label{eq:MassFromHarmonicEstimator}
\end{equation}
where the empirical parameters $K=2.684 \pm 0.001\MeV\unit{cm}^{-1}$
and $C=3.375 \pm 0.001\MeV\unit{cm}^{-1}$ are determined from
data using a sample of low-momentum protons.
As the momentum reconstruction is done assuming $\abs{Q}=1e$ particles,
Eq.~(\ref{eq:MassFromHarmonicEstimator}) leads to
an accurate mass  reconstruction only for singly charged particles.

The HSCP candidates are preselected using the \ias discriminator
because it has a better signal-to-background discriminating power compared to the \ih estimator or the mass.
Nonetheless, the mass is used at the last stage of the analysis, after the \ias selection,
to further discriminate between signal and backgrounds since the latter tend to have a low reconstructed mass.

\vspace{1cm}
\subsection{Time of flight measurements}

The time of flight to the muon system can be used to discriminate between
particles travelling at near the speed of light and slower candidates.
Both the DT and the CSC muon systems measure the time of each hit.
In the DT, the precision position is obtained from this time measurement.
The synchronization works in a such a way that a relativistic muon produced at the interaction point gives
an aligned pattern of hits in consecutive DT layers.
For a slower HSCP particle, hits in each DT layer will be reconstructed as shifted with
respect to its true position and will form a zigzag pattern with an offset proportional to the
particle delay, $\delta_t$. In the CSC the delay is measured for each hit separately.
Each $\delta_t$ measurement can be used to determine the track
$\beta$ via the equation:
\begin{equation}
 \beta^{-1}= 1+ \frac{c \delta_t}{L}
 \label{betatotof}
\end{equation}
where $L$ is the flight distance.
The track $\beta^{-1}$ value is calculated as the weighted
average of the  $\beta^{-1}$ measurements from the DT and CSC systems associated with the track.
The weight for the $i^{\mathrm{th}}$ DT measurement is given by
\begin{equation}
 w_{i} = \frac{(n-2)}{n}\frac{L_{i}^{2}}{\sigma_{\mathrm{DT}}^{2}}
\end{equation}
where $n$ is the number of $\phi$ projection measurements found in the
muon chamber producing the measurement
and $\sigma_{\mathrm{DT}}$ is the time resolution of the DT
measurements, for which the measured value of 3\unit{ns} is used.
The factor $(n-2)/n$ accounts for  residuals computed
using the two parameters of a straight line determined from the same
$n$ measurements. The minimum number of hits in a given DT chamber that allows
for at least one residual calculation is $n=3$.
The weight for the $i$th CSC measurement is given by
\begin{equation}
 w_{i} =\frac{L_{i}^{2}}{\sigma_{i}^{2}}
\end{equation}
where $\sigma_{i}$, the measured time resolution, is 7.0\unit{ns}
for cathode strip measurements and 8.6\unit{ns} for anode wire measurements.

The resolution on the weighted average $\beta^{-1}$ measurement is approximately 0.065 in both the DT and CSC subsystems.

\section{Data selection \label{sec:presel}}

All events pass a trigger requiring either a reconstructed muon with high
transverse momentum or large \MET,
calculated using an online particle-flow algorithm~\cite{CMS-PAS-JME-10-003,ref:PAS-PFT-09-001,CMS-PAS-PFT-10-001}.

The muon trigger is more efficient than the \MET trigger for all HSCP models considered with the exception
of the charge-suppressed $R$-hadron model,
but it is not efficient for particles that are slow ($\beta<0.6$).

The \MET trigger can recover some events in which the HSCP
is charged in the tracker and neutral in the muon subsystem.
The particle-flow algorithm rejects tracks reconstructed only in
the tracker and having a track \pt significantly greater than the matched energy
deposited in the calorimeter~\cite{ref:PAS-PFT-09-001},
as would be the case for HSCPs that become neutral in the calorimeter.
Thus only an HSCP's energy deposit in the calorimeter, roughly 10--20\GeV, will be included in the \MET\ calculation.
Where one or more HSCPs fail to be reconstructed as
muon candidates, the events may appear to have significant \MET.

For both the \tkonly and the \tktof analyses, the muon high-level trigger
requires a muon candidate with $\pt > 50$\GeV
and the \MET\ trigger requires
$\MET > 170$\GeV.
Using these two triggers for both analyses allows for increased sensitivity
to HSCP candidates that arrive in the muon system very late, as well as for
hadronlike HSCPs, which may be charged only in the tracker.

Offline, for the \tkonly analysis, all events are required
to have a candidate track with
$\pt > 55$\GeV as measured in the tracker, relative
uncertainty in \pt ($\sigma_{\pt}/\pt$) less than 0.25,
$\abs{\eta} < 2.1$,  and the track fit $\chi^2/\mathrm{dof} < 5$. The magnitudes of the impact
parameters $d_z$ and $d_{xy}$ must both be less than 0.5\unit{cm}, where $d_z$ and $d_{xy}$
are the longitudinal and transverse impact parameters
with respect to the vertex with the smallest $d_z$.
The requirements on the impact parameters are very loose compared to the resolutions for tracks in the tracker.
Candidates must pass isolation requirements in the tracker and
calorimeter.  The tracker isolation criterion is
$\sum\pt < 50$\GeV, where the sum is over all tracks (except the candidate)
within $\Delta R = \sqrt{\smash[b]{(\Delta \eta)^2 + (\Delta \phi)^2}} = 0.3$ of the candidate track.
The calorimeter isolation criterion is $E/p = 0.3$, where $E$
is the sum of energy deposited in the calorimeter towers
within $\Delta R = 0.3$ and $p$ is the track momentum reconstructed
from the tracker.
Candidate tracks must have at least two measurements in the silicon
pixel detector and
at least six measurements in the strip detectors.
In addition, there must be measurements in at least 80\% of
the silicon layers between the first and last measurements of
the track.  To reduce the contamination from clusters
with a large energy deposition due to overlapping tracks,
a~filtering procedure is applied to remove clusters in the
silicon strip tracker that are not
consistent with the passage of a singly charged particle (i.e.,
a narrow cluster with most of the energy deposited in one or
two strips).
After cluster filtering, there must be at least six measurements
in the silicon tracker that are used for the \dedx calculation.

The \tktof analysis applies the same criteria, but
additionally requires a reconstructed muon matched to the track
in the inner detectors.  At least eight independent time measurements
are needed for the TOF computation.  Finally, $1/\beta > 1$
and $\sigma_{1/\beta} < 0.15$ are required.

\section{Background estimation \label{sec:bkgpred}}

For background estimation we follow the procedure described in our previous work~\cite{Chatrchyan:2013oca}.
Candidates passing the preselection
(Section~\ref{sec:presel})
are subject to either two or three additional criteria to improve
the signal-to-background discrimination.
By choosing two uncorrelated criteria it is possible to predict
the background using the ABCD (matrix) method. In this approach, the expected background
 in the signal region, $D$, is estimated by $BC/A$, where $B$ and $C$ are
the number of candidates that fail the first or second criterion,
respectively, while $A$ is the number of candidates that fail both criteria.

Results  are based upon a comparison of the number of candidates passing the
 selection criteria defining the signal region
 with the number of predicted background events in that region.
Fixed selections on the appropriate set of \ias, \pt, and \invbeta are used to define the final signal
region (and the regions for the background prediction).
The values are chosen to give the best discovery potential over
the signal mass regions of interest.

For the \tkonly analysis, the two criteria are
$\pt > 65$\GeV and \ias $>$0.3.
The candidates passing only the \ias requirement fall into the $B$ region and those passing only the \pt requirement fall into the $C$ region.
The $B$ and $C$ candidates are then used to form binned probability
density functions in \ih and $p$, respectively, such that, using the mass
value (Eq.~(\ref{eq:MassFromHarmonicEstimator})),
the full mass spectrum of the background in the
signal region $D$ can be predicted.  However, the $\eta$
distribution of candidates with low \dedx differs from the
distribution of candidates with high \dedx.
To correct for this,
events in the $C$ region are weighted
such that the $\eta$ distribution matches that in the $B$ region.

For the \tktof analysis, a three-dimensional matrix method is used with
$\pt > 65$\GeV, $\ias > 0.175$, and $\invbeta > 1.25$, creating eight regions labeled $A$--$H$.
 Region $D$ represents the signal region, with events passing all three criteria.
The candidates in the $A$, $F$, and $G$ regions pass only the \invbeta, \ias, and \pt criteria, respectively,
while the candidates in the $B$, $C$, and $H$ regions fail only the \pt, \ias, and \invbeta
criteria, respectively.  The $E$ region contains events that fail all three criteria.
Background estimates can be made from several different combinations
of these regions.  The combination $D = AGF/E^2$ is used because it yields
the smallest statistical uncertainty.  As in the \tkonly analysis, events in the $G$ region are reweighted to match the
$\eta$ distribution in the $B$ region.  The spread in
background estimated from the other combinations is less than 20\%, which is taken as the systematic
uncertainty in the collision background estimate.  The same 20\% systematic uncertainty is used for the \tkonly analysis.

\begin{figure}
 \centering
  \includegraphics[width=0.48\textwidth]{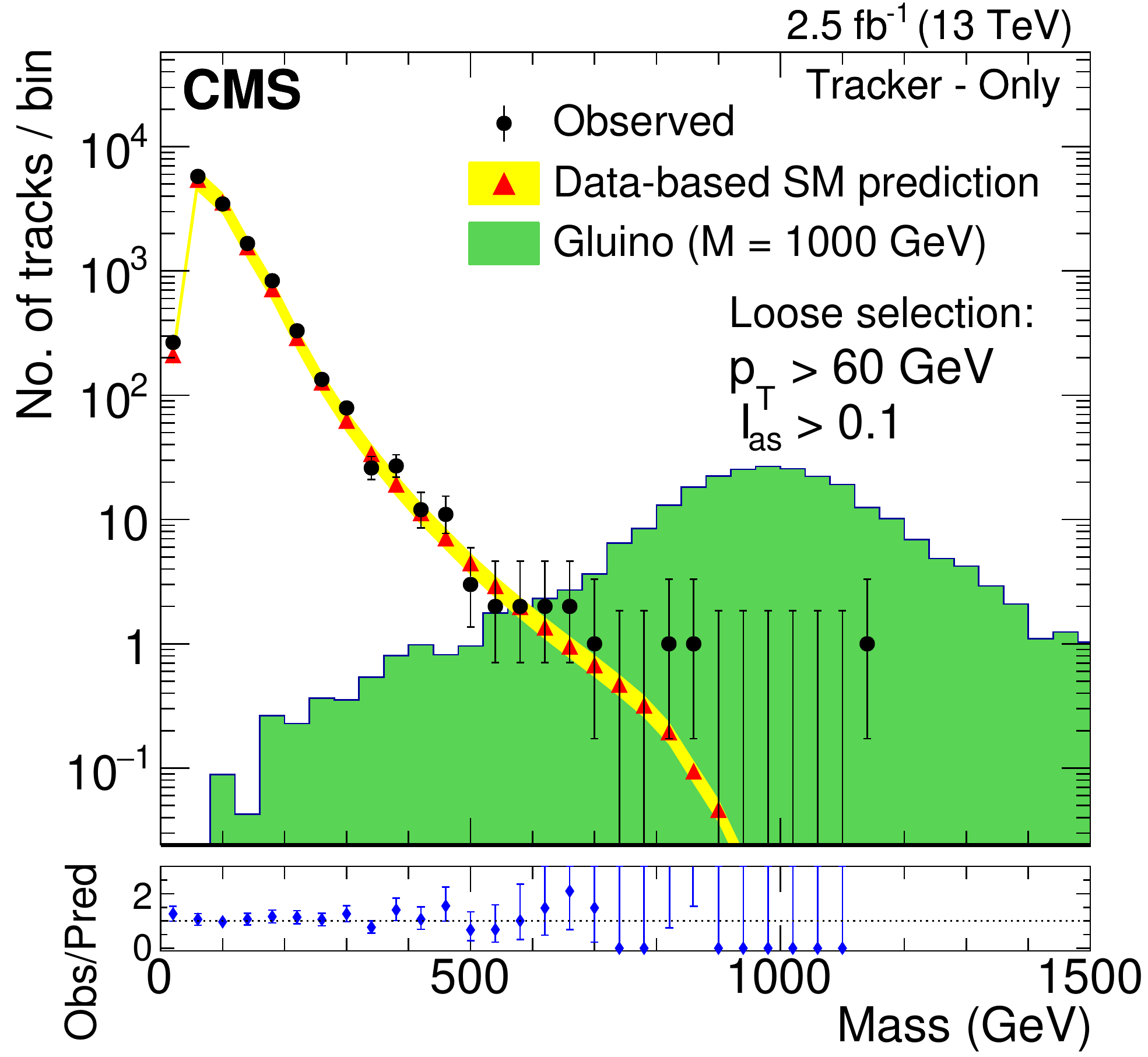}
  \includegraphics[width=0.48\textwidth]{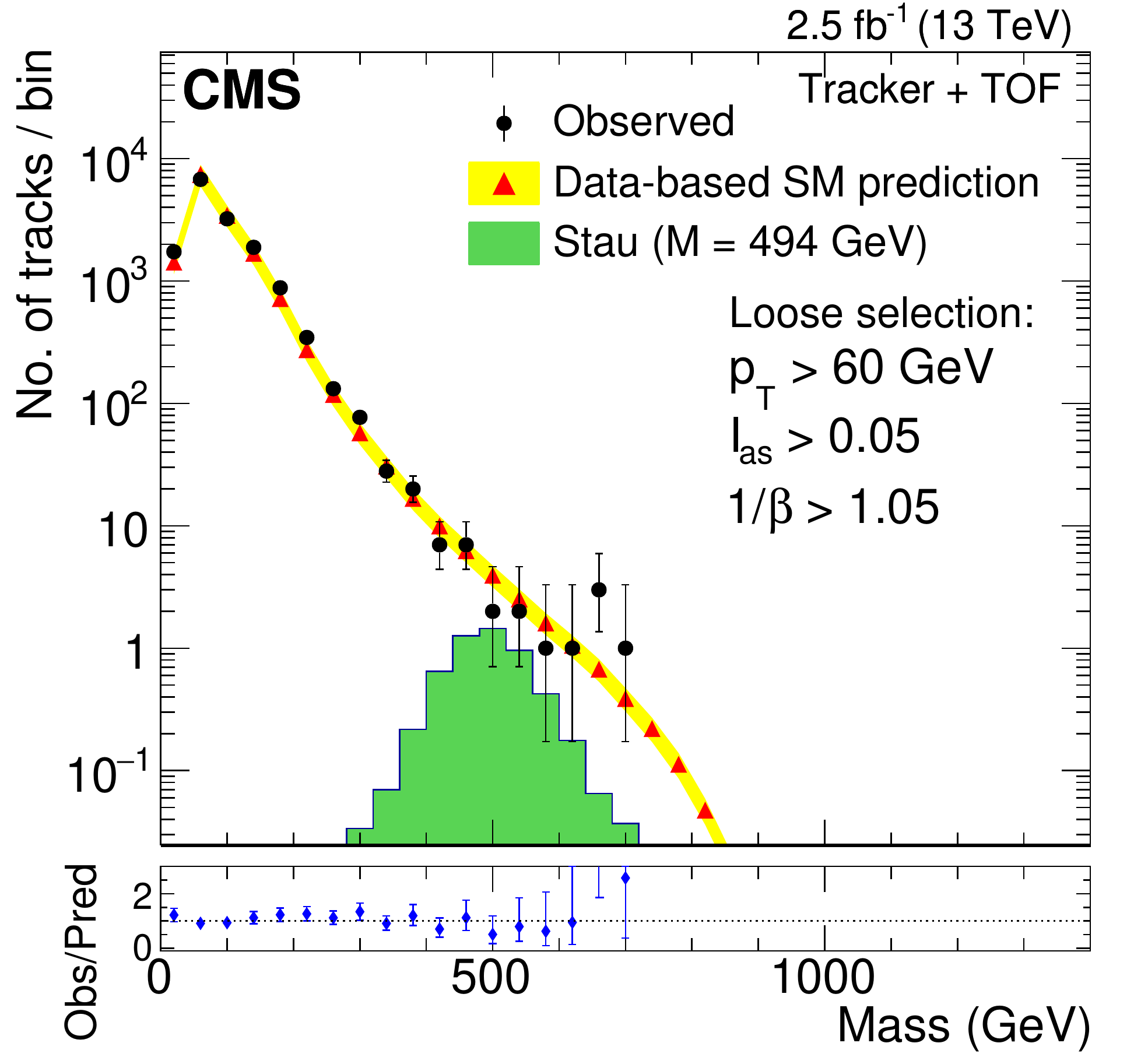}
 \caption{Observed and predicted mass spectra for loose selection candidates
    in the \tkonly (\cmsLeft) and \tktof (\cmsRight) analyses.
    The expected distributions for representative signals are shown as histograms.
    \label{fig:LooseMassDistribution}}
\end{figure}

In order to check the background prediction, samples with a
loose selection,
which would be dominated by background tracks, are used for the \tkonly and \tktof analyses.
The loose selection sample for the \tkonly analysis is defined
as $\pt>60$\GeV and $I_{as}>0.10$.  The loose selection
sample for the \tktof analysis is defined by
$\pt>60$\GeV, $I_{as}>0.05$, and $1/\beta>1.05$.
Figure~\ref{fig:LooseMassDistribution} shows the observed and estimated
mass spectra for these samples.

For both analyses, an additional requirement on the reconstructed
mass is applied.
The specific requirement is adapted to each HSCP model.
For a given signal mass and model, the mass requirement is
$M \geq M_{\text{reco}} - 2\sigma$, where $M_{\text{reco}}$ is
the average reconstructed mass for the given mass $M_{\mathrm{HSCP}}$ and
$\sigma$ is the expected resolution.  Simulation is used to
determine $M_\text{reco}$ and $\sigma$.

\begin{figure}
\centering
 \includegraphics[width=0.48\textwidth]{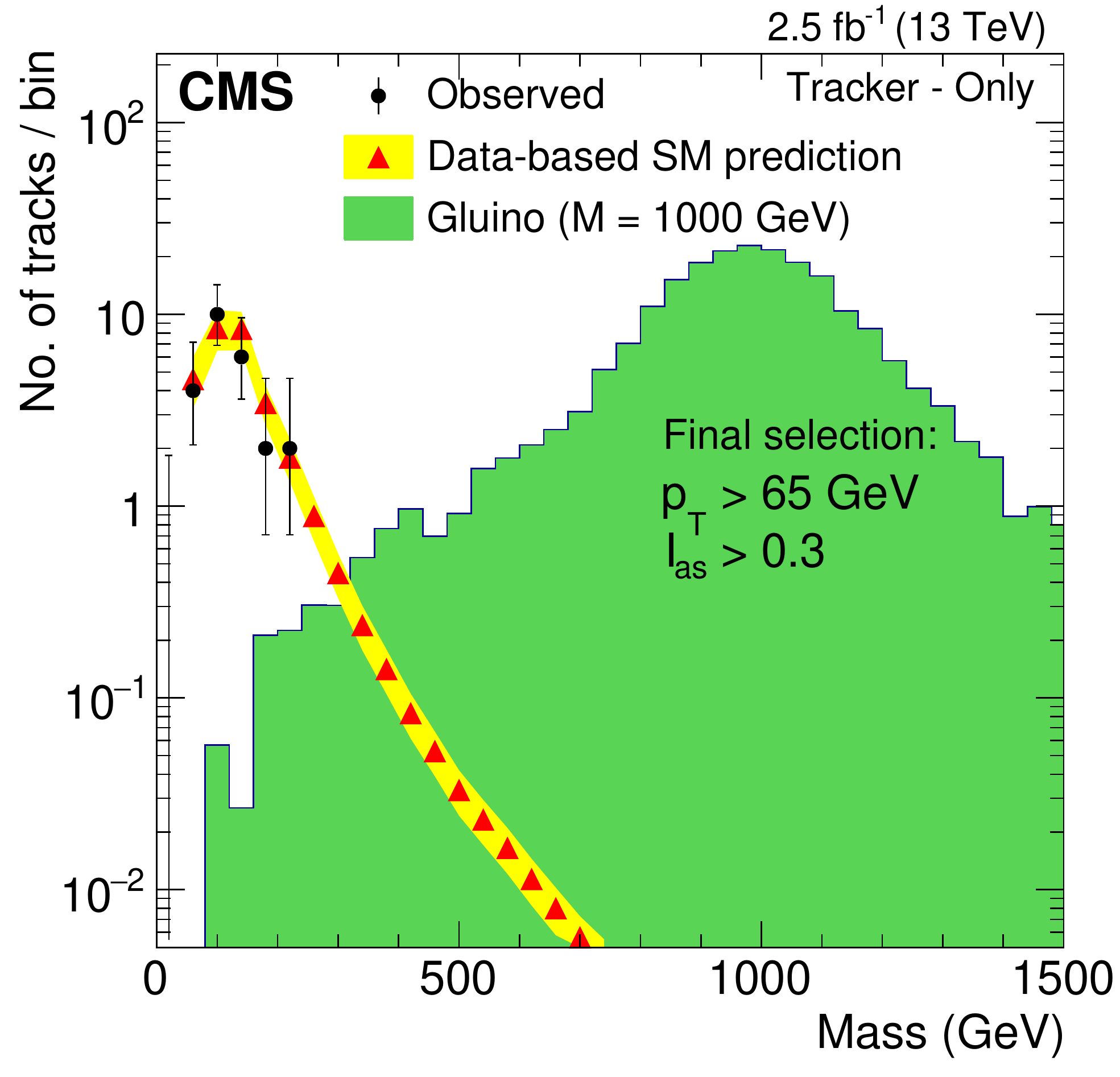}
 \includegraphics[width=0.48\textwidth]{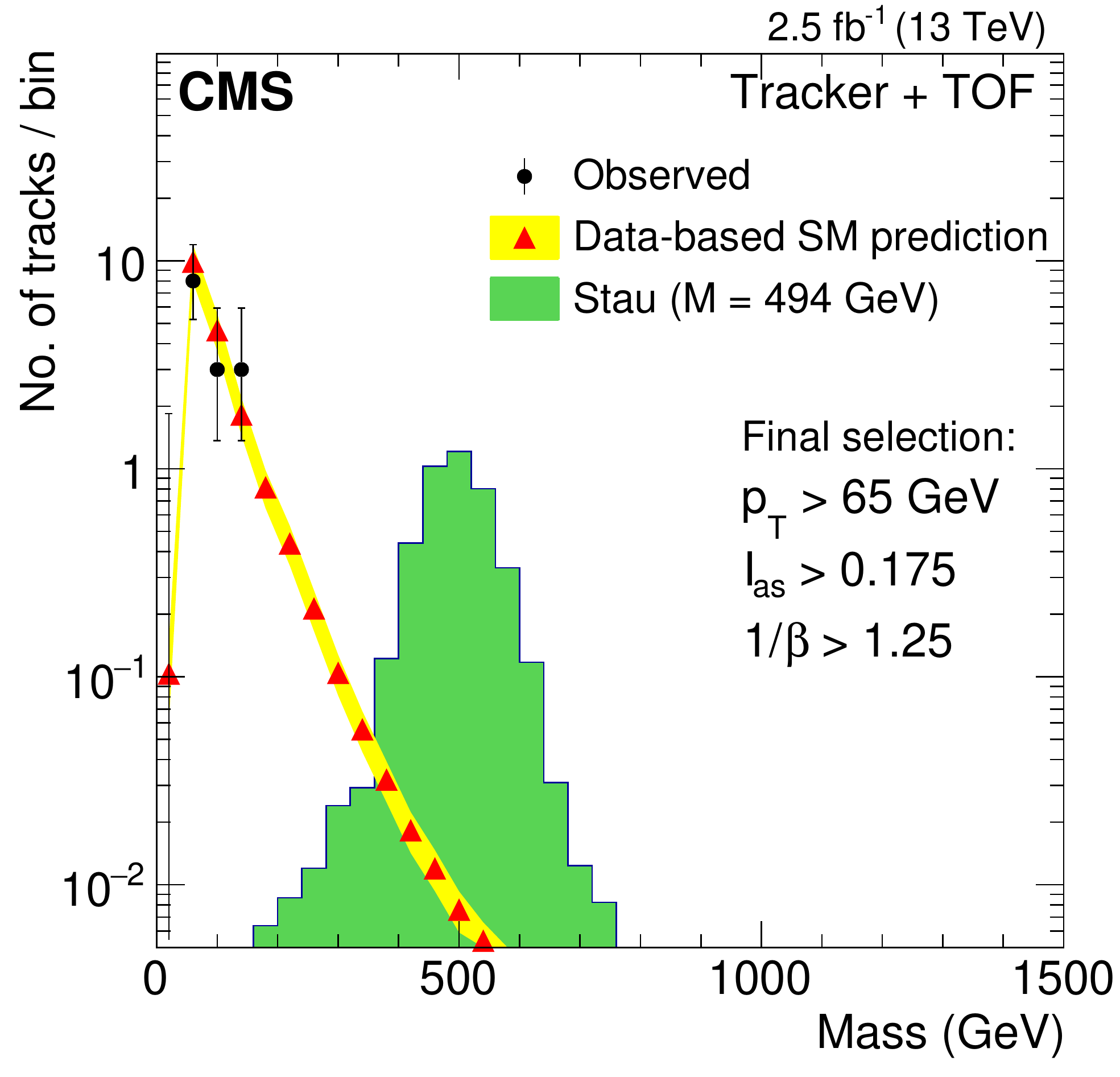} \\
 \caption{Observed and predicted mass spectra for candidates passing the final selection in
    the \tkonly (\cmsLeft) and \tktof (\cmsRight) analyses.
    The expected distributions for representative signals are shown as histograms.
    \label{fig:TightMassDistribution}}
\end{figure}

Table~\ref{tab:finalsel} lists the final selection criteria, the predicted
number of background events, and the number of events observed in
the signal region.  Agreement between prediction and observation
is seen for both  the \tkonly and the
\tktof analyses. Figure~\ref{fig:TightMassDistribution} shows the predicted and observed  mass
distributions for the \tkonly and the \tktof analyses with the final selection.

\begin{table*}
\topcaption{
Selection criteria for the two analyses with the number of predicted and observed
events. In the background prediction, the statistical and systematic uncertainties are added
in quadrature.
\label{tab:finalsel}}
\newcolumntype{X}{D{,}{\,\pm\,}{3,3}}
\centering
\begin{scotch}{lcccc{c}@{\hspace*{5pt}}Xr}
                             &\multicolumn{4}{c}{Selection requirements}&& \multicolumn{2}{c}{Numbers of events} \\
                             &&&& && \multicolumn{2}{c}{$\sqrt{s}=13$\TeV} \\ \cline{2-5}\cline{7-8}\\[-2.2ex]
                             & \pt                  & \multicolumn{1}{c}{$I_{as}$}  & \multicolumn{1}{c}{$1/\beta$} & Mass    && \multicolumn{1}{c}{Pred.} & \multicolumn{1}{r}{Obs.}   \\
                             & (\GeVns{})                   &                            &                            & (\GeVns{})   &&                &     \\ \hline
   \multirow{4}{*}{\tkonly} & \multirow{4}{*}{$>$65} & \multirow{4}{*}{$>$0.3}   & \multirow{4}{*}{\NA}             & $>$0 && 28.7,6.0    & 24  \\
                             &                         &                            &                            & $>$100 && 20.7,4.4    & 15  \\
                             &                         &                            &                            & $>$200 && 3.8,0.8  & 2  \\
                             &                         &                            &                            & $>$300 && 0.82,0.18  & 0  \\
                             &                         &                            &                            & $>$400 && 0.25,0.05 & 0  \\ \hline
   \multirow{4}{*}{\tktof}  & \multirow{4}{*}{$>$65} & \multirow{4}{*}{$>$0.175} & \multirow{4}{*}{$>$1.250}       & $>$0 && 18.2,3.7    & 14  \\
                             &                         &                            &                            & $>$100 && 5.4,1.1    & 4   \\
                             &                         &                            &                            & $>$200 && 0.90,0.19  & 0   \\
                             &                         &                            &                            & $>$300 && 0.06,0.04  & 0  \\
  \end{scotch}
\end{table*}

\section{Systematic uncertainties}

The sources of systematic uncertainty considered are
those related to the background prediction, the signal acceptance, and  the integrated luminosity.
The uncertainty in the integrated luminosity is 2.7\% at $\sqrt{s} = 13$\TeV~\cite{CMS:2016eto}.
   The uncertainties in the collision background predictions are estimated to be at the level
of 20\% for the \tkonly and the \tktof analyses, as described in Section~\ref{sec:bkgpred}.

The signal acceptance is obtained from MC
samples of the various signals processed through
the full detector simulation (Section~\ref{sec:signals}).
Systematic uncertainties are derived by
comparing the response of the detector in the data and simulation.
The relevant sources of uncertainty  are discussed below.

The signal trigger efficiency is dominated by the muon triggers
efficiency, for all the models except the charge-suppressed ones.  The uncertainty in the muon
trigger efficiency
has many contributions.
It is estimated from the difference between the trigger efficiency in data and that seen in
simulation, using $\Z(\mu\mu)$ data.
For genuine muons, the trigger efficiency uncertainty is 3\%.

 For slow moving particles, the effect of the timing synchronization of the muon system is tested by shifting
the arrival times in simulation by the synchronization
accuracy observed in data, resulting in an efficiency change of less than 4\%
for most samples but up to 8\% for the 2.4\TeV gluino sample.
The uncertainty in the \MET trigger efficiency is found by
varying the jet energy scale in the simulation of the high-level trigger by its uncertainty in data.
The \MET uncertainty is found
to be less than 12\% for all samples.
The total trigger uncertainty is found to be less than 13\% for
all the samples, since the muon trigger inefficiencies are often
compensated by the \MET trigger and vice versa.

Low-momentum protons are used to compare the observed and simulated distributions of \ih and \ias\,
that reflect the energy loss in the silicon tracker.
The \dedx distributions of signal samples
are varied by the observed differences in order to estimate the systematic uncertainty.
The uncertainty in the signal acceptance is usually less than 10\%, and is at most 15\%.

{\tolerance=800 
Bias in the energy loss measurement due to highly ionizing particles (HIP),
such as low-momentum protons
produced in pp collisions earlier than the triggering collision,
was also considered as a~source of uncertainty in the \ih estimate.
In 2015, the LHC collision frequency was doubled,
with bunches colliding every 25\unit{ns}
compared to  collisions every 50\unit{ns} in 2012, causing an increase of the  HIP rate.
The contribution of HIPs was included in simulations with the rate observed during the 2015 data taking.
The uncertainty in this rate is found to be  25\% and 80\% for pixel and strip sensors, respectively.
Varying the HIP rate in the simulation by these amounts leads to a change in signal acceptance of at most 4\% for both analyses.
\par}

Dimuon events are used to test the MC
simulation of \invbeta by comparing with data.
An offset of at most 1.5\% is found for the muon system.
The resulting uncertainty (labeled ``Time of flight'' in Table~\ref{tab:systuncertainties})
in the signal acceptance is found to be less than 5\% by shifting \invbeta\ by this amount.

As in Ref.~\cite{Chatrchyan:2012sp}, the uncertainties
in the efficiencies for muon~\cite{MUO-10-004}
and track~\cite{CMS-PAS-TRK-10-002}  reconstruction are each less than 2\%.
The track momentum uncertainty is estimated by shifting the momentum of the inner track,
 as in Ref.~\cite{Chatrchyan:2012sp}.  This uncertainty is found to be less than 5\%
for most of the samples, increasing to 20\% for masses above 2\TeV.

The uncertainty in the number of pileup events
is evaluated by varying  $\pm$5\% the minimum bias cross section used to
calculate the weights applied to signal events in  order  to reproduce
the pileup observed in data.
The uncertainties due to pileup estimated with this
procedure are less than 1\%.

The total systematic uncertainty in the signal acceptance is the sum
in quadrature of the uncertainties due to the sources discussed above.
For almost all signal models, it is less than 20\% for both analyses.
Only for the \tktof analysis of the gluino ($f=0.5$) sample it is larger,
but does not exceed 25\%.

Table~\ref{tab:systuncertainties}
summarizes the systematic uncertainties for the two analyses.
As the uncertainty often depends on the model and HSCP mass, the largest systematic uncertainty is reported for each source.

\begin{table}
 \topcaption{Systematic uncertainties for the two HSCP searches.  All
   values are relative uncertainties in the signal acceptance for the \tkonly and \tktof analyses.
   \label{tab:systuncertainties}}
 \centering
  \begin{scotch}{ l  r  r }
Source of systematic uncertainty  & \multicolumn{2}{c}{Relative uncertainty (\%)}  \\ \hline
Signal acceptance                   & \multicolumn{1}{c}{\tkonly} & \multicolumn{1}{c}{\tktof}  \\
~~-~~Trigger efficiency             & $13$ & $13$  \\
~~-~~Track momentum scale           & $<$20 & $<$20  \\
~~-~~Track reconstruction           & $<$2 & $<$2 \\
~~-~~Ionization energy loss         & $<$15 & $<$15  \\
~~-~~HIP background effect          & $<$3 & $<$4  \\
~~-~~Time of flight                 & \NA & $<$5  \\
~~-~~Muon reconstruction            & \NA & 2   \\
~~-~~Pileup                        & $<$1 & $<$1  \\
Tot. uncert. in signal acceptance & $<$20 & $<$25 \\
Collision background uncert.              &  20    &  20   \\
Luminosity uncertainty               & 2.7& 2.7 \\
  \end{scotch}
\end{table}

\section{Results \label{sec:results}}

\begin{figure}
 \centering
  \includegraphics[width=0.48\textwidth]{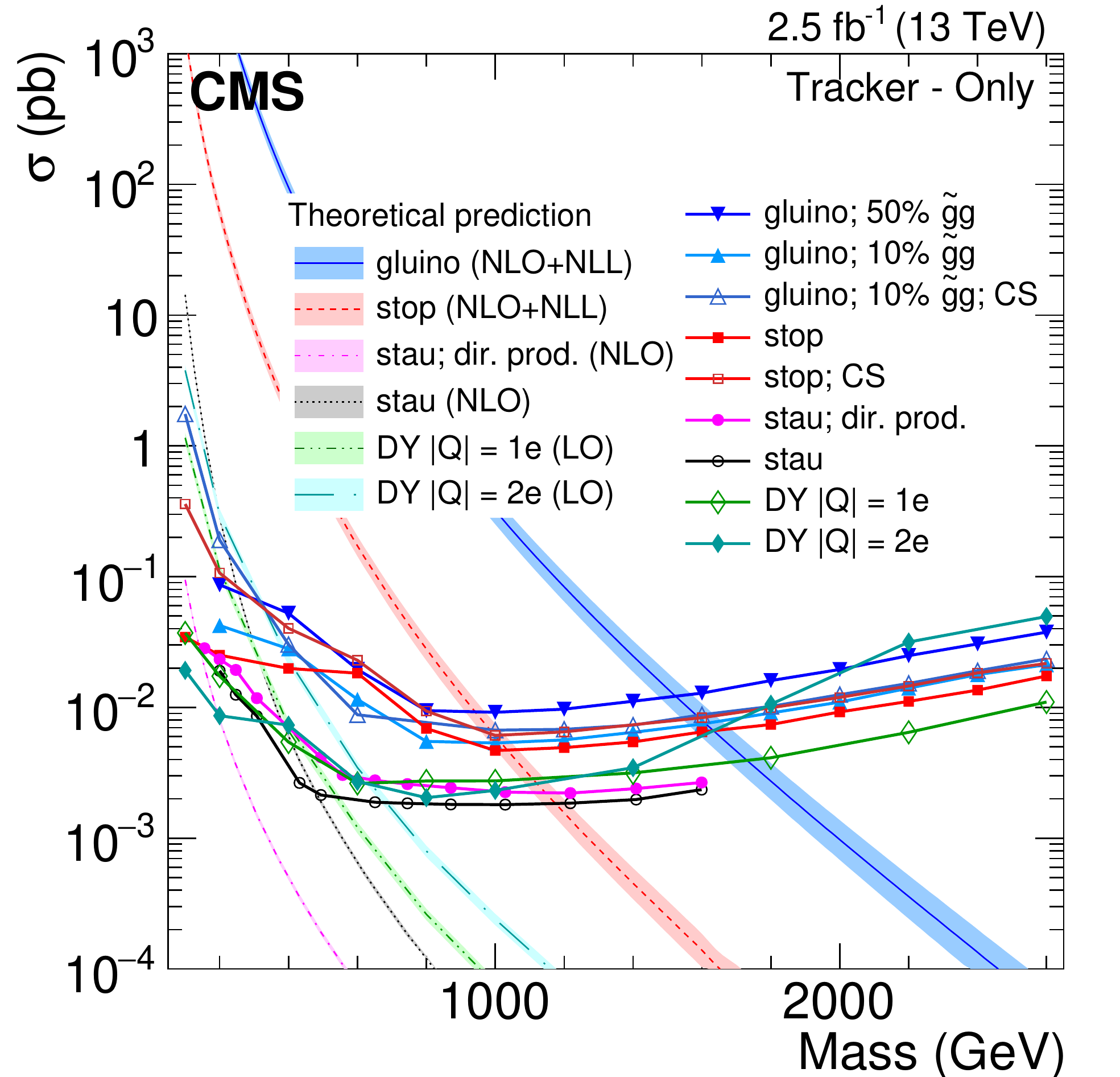}
  \includegraphics[width=0.48\textwidth]{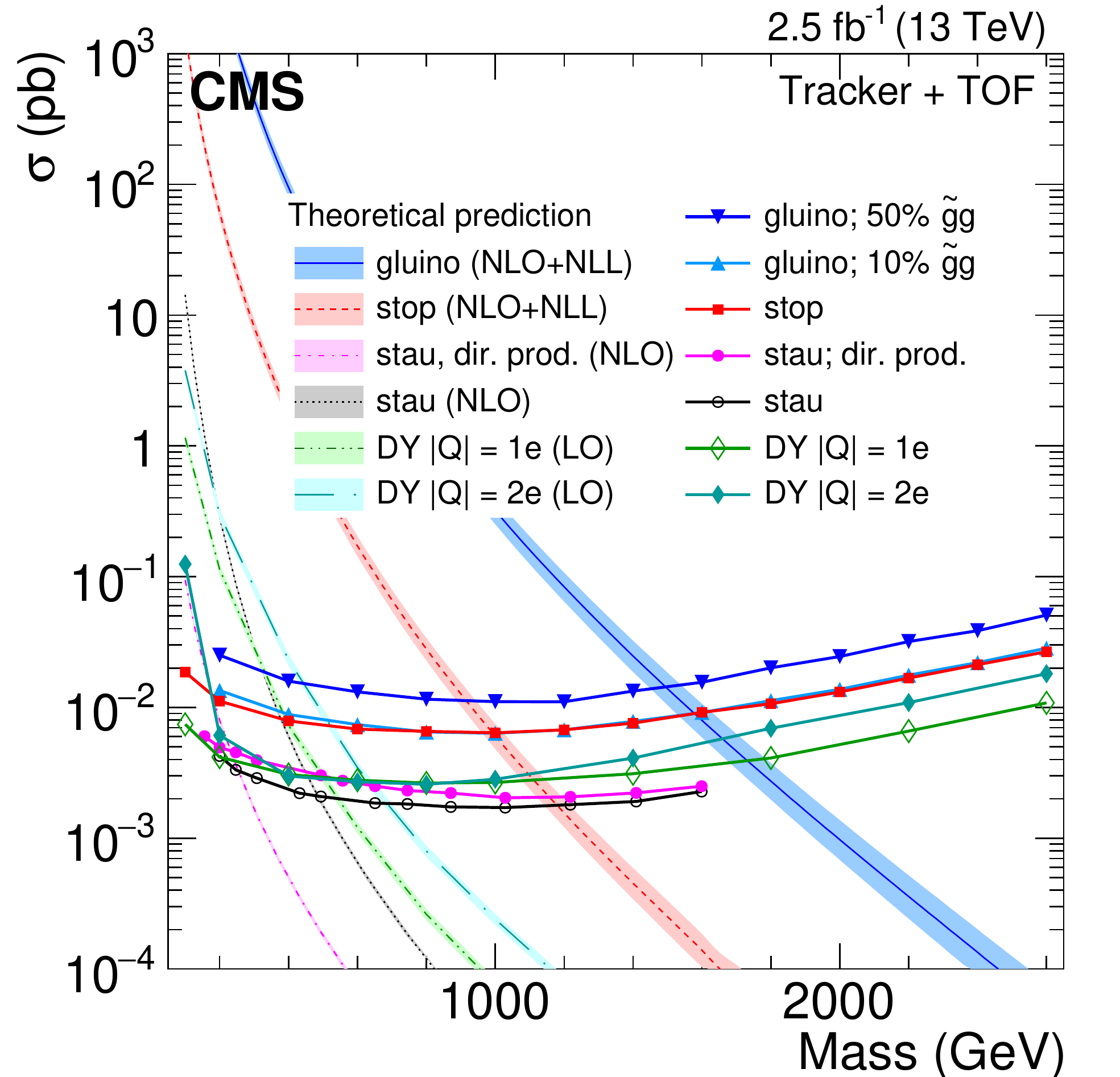}
 \caption{Results of the HSCP search as the cross section upper limits at 95\% \CL for various signal models for the
   \tkonly analysis (\cmsLeft) and \tktof    analysis (\cmsRight) at $\sqrt{s} = 13$\TeV.
   In the legend, ``CS'' stands for charge-suppressed interaction model.
   \label{fig:limits1}}
\end{figure}

No significant excess of events is observed above the predicted background.
Cross section limits are placed at 95\% \CL using a
CL$_\mathrm{s}$ approach~\cite{Junk:1999kv,READ:JPG2002,CMS-NOTE-2011-005} where
a profile likelihood technique~\cite{Cowan:2010js} is used. It utilizes a log-normal
model~\cite{Eadie, James} for the nuisance parameters, which are
the integrated luminosity, the signal acceptance, and the
expected background in the signal region.
The observed limits are shown in Fig.~\ref{fig:limits1} for both
the \tkonly and the \tktof analyses along with the theoretical predictions.
The theoretical cross sections are computed
at NLO or NLO+NLL~\cite{Kulesza:2008jb,Kulesza:2009kq,Beenakker:2009ha,Beenakker:2010nq} using
\PROSPINO~\cite{Beenakker:1996ed} with CTEQ6.6M PDFs~\cite{Nadolsky:2008zw}.
The uncertainty bands of the theoretical cross sections
include the PDF uncertainty, the renormalization and factorization scale uncertainties,
and the uncertainty in $\alpha_{s}$.
The 95\% \CL limits on the production cross sections
are shown in Tables~\ref{tab:limitsGluino},~\ref{tab:limitsStop},~\ref{tab:limitsStau}, and \ref{tab:limitsDY}
for long-lived gluino, top squark, tau slepton, and modified Drell--Yan signals, respectively.
The limits were determined from the numbers of events passing
all final criteria (including the mass criteria).

Mass limits are obtained from the intersection of the observed
limit and the central value of the theoretical cross section.
The \tkonly analysis excludes $f=0.1$ gluino masses below 1610 (1580)\GeV
for the cloud interaction model (charge-suppressed model).
Top squark masses below 1040\,(1000)\GeV are excluded for the  cloud (charge-suppressed) models.
In addition, the \tktof analysis excludes \stau masses below
490\,(240)\GeV for the GMSB (direct pair production) model. Drell--Yan signals with
$\abs{Q} = 1e$ ($2e$) are excluded below 550\,(680)\GeV.

The mass limits obtained at  $\sqrt{s}=13$\TeV for various HSCP signal
models are summarized in Table \ref{tab:MassLimits}
and compared with earlier results at  $\sqrt{s}=7$ and 8\TeV~\cite{Chatrchyan:2013oca}.
A significant increase in mass limit is obtained for all models with
large QCD production cross section (gluinos, top squarks,
and inclusive production of GMSB tau sleptons), arising from the higher
center-of-mass energy pp collisions delivered by the LHC.
For scenarios with much smaller cross-sections, directly pair-produced tau sleptons
and Drell--Yan signals with $\abs{Q} = 1e$, the results do not improve, because
the larger integrated luminosity at 7 and 8\TeV with respect to that at 13\TeV
prevails over the effect of the increase of the centre-of-mass energy.
For the $\abs{Q} = 2e$ analysis, results from the previous analysis optimized
for multiply charged signals~\cite{Chatrchyan:2013oca} are also provided.

\section{Summary}
A search for heavy stable charged particles produced in proton-proton collisions
at $\sqrt{s}=13$\TeV using the CMS detector is presented.
Two complementary analyses were
performed: using only the tracker and using
both the tracker and the muon system.
The data are found to be compatible with the expected background. Mass limits for long-lived gluinos,
top squarks, tau sleptons, and multiply charged particles are calculated.
The models for $R$-hadronlike HSCPs include a varying fraction of
\PSg-gluon hadronization
and two different interaction models leading to a variety of exotic experimental signatures.
The limits  are significantly improved over those from
 Run~1 of the LHC, and the limits on long-lived
gluinos, ranging up to 1610\GeV,  are the most stringent to date.

\begin{table*}
 \topcaption{Summary of the search for long-lived gluinos: the \pt (\GeVns{}),  \ias, \invbeta, and mass thresholds M\,(\GeVns{}) requirements,
the predicted and observed yields passing these criteria, and the resulting
expected (exp.) and observed (obs.) cross section limits.
The signal efficiencies and theoretical (theo.) cross sections are also listed.
  \label{tab:limitsGluino}}
 \centering
 \cmsTableResize{
 \begin{scotch}{|c|cccc|xc|c|lll|}
 Mass & \multicolumn{4}{c|}{Requirements} & \multicolumn{2}{c|}{Yields} & Signal & \multicolumn{3}{c|}{$\sigma$ (pb)}  \\
 (\GeVns{})& \pt   & \ias    & \invbeta  & M  & \multicolumn{1}{c}{SM predicted} & data & eff.    & theo. & exp. & obs.   \\ \hline
 \multicolumn{11}{|c|}{Gluino ($f=0.1$)  with the \tkonly analysis}  \\ \hline
  400 &     65 & 0.3 & \NA &   60 & 28.000,5.880 & 23 & 0.167 & 9.5$\times 10^{+1}$ & 3.7$\times 10^{-2}$ & 2.8$\times 10^{-2}$ \\
  800 &     65 & 0.3 & \NA &  350 &  0.435,0.093 &  0 & 0.223 & 1.5               & 5.5$\times 10^{-3}$ & 5.5$\times 10^{-3}$ \\
 1200 &     65 & 0.3 & \NA &  590 &  0.046,0.010 &  0 & 0.220 & 8.4$\times 10^{-2}$ & 5.6$\times 10^{-3}$ & 5.6$\times 10^{-3}$ \\
 1600 &     65 & 0.3 & \NA &  720 &  0.017,0.004 &  0 & 0.166 & 8.0$\times 10^{-3}$ & 7.5$\times 10^{-3}$ & 7.5$\times 10^{-3}$ \\
 2000 &     65 & 0.3 & \NA &  770 &  0.012,0.003 &  0 & 0.112 & 9.7$\times 10^{-4}$ & 1.1$\times 10^{-2}$ & 1.1$\times 10^{-2}$ \\
 2400 &     65 & 0.3 & \NA &  800 &  0.012,0.002 &  0 & 0.072 & 1.3$\times 10^{-4}$ & 1.8$\times 10^{-2}$ & 1.8$\times 10^{-2}$ \\
\hline
 \multicolumn{11}{|c|}{Gluino charge-suppressed ($f=0.1$)  with the \tkonly analysis} \\ \hline
  400 &     65 & 0.3 & \NA &  120 & 15.600,3.300 & 10 & 0.092 & 9.5$\times 10^{+1}$ & 4.9$\times 10^{-2}$ & 3.0$\times 10^{-2}$ \\
  600 &     65 & 0.3 & \NA &  250 &  1.690,0.369 &  0 & 0.141 & 9.1               & 1.2$\times 10^{-2}$ & 8.8$\times 10^{-3}$ \\
 1200 &     65 & 0.3 & \NA &  580 &  0.050,0.011 &  0 & 0.183 & 8.4$\times 10^{-2}$ & 6.8$\times 10^{-3}$ & 6.8$\times 10^{-3}$ \\
 1600 &     65 & 0.3 & \NA &  680 &  0.023,0.005 &  0 & 0.142 & 8.0$\times 10^{-3}$ & 8.8$\times 10^{-3}$ & 8.8$\times 10^{-3}$ \\
 2000 &     65 & 0.3 & \NA &  670 &  0.024,0.005 &  0 & 0.099 & 9.7$\times 10^{-4}$ & 1.3$\times 10^{-2}$ & 1.3$\times 10^{-2}$ \\
 2400 &     65 & 0.3 & \NA &  680 &  0.023,0.005 &  0 & 0.066 & 1.3$\times 10^{-4}$ & 1.9$\times 10^{-2}$ & 1.9$\times 10^{-2}$ \\
\hline
 \multicolumn{11}{|c|}{Gluino ($f=0.5$)  with the \tkonly analysis} \\ \hline
  400 &     65 & 0.3 & \NA &   50 & 28.700,6.030 & 24 & 0.094 & 9.5$\times 10^{+1}$ & 6.6$\times 10^{-2}$ & 5.2$\times 10^{-2}$ \\
  800 &     65 & 0.3 & \NA &  340 &  0.491,0.105 &  0 & 0.129 & 1.5               & 9.5$\times 10^{-3}$ & 9.5$\times 10^{-3}$ \\
 1200 &     65 & 0.3 & \NA &  580 &  0.050,0.011 &  0 & 0.127 & 8.4$\times 10^{-2}$ & 9.7$\times 10^{-3}$ & 9.7$\times 10^{-3}$ \\
 1600 &     65 & 0.3 & \NA &  710 &  0.018,0.004 &  0 & 0.096 & 8.0$\times 10^{-3}$ & 1.3$\times 10^{-2}$ & 1.3$\times 10^{-2}$ \\
 2000 &     65 & 0.3 & \NA &  760 &  0.013,0.003 &  0 & 0.063 & 9.7$\times 10^{-4}$ & 2.0$\times 10^{-2}$ & 2.0$\times 10^{-2}$ \\
 2400 &     65 & 0.3 & \NA &  740 &  0.014,0.003 &  0 & 0.040 & 1.3$\times 10^{-4}$ & 3.1$\times 10^{-2}$ & 3.1$\times 10^{-2}$ \\
 \end{scotch}
 }
\end{table*}

\begin{table*}
 \topcaption{
Summary of the search for long-lived top squarks:  the \pt\,(\GeVns{}),  \ias, \invbeta, and mass thresholds M\,(\GeVns{}) requirements,
the predicted and observed yields passing these criteria, and the resulting
expected (exp.) and observed (obs.) cross section limits.
The signal efficiencies and theoretical (theo.) cross sections are also listed.
  \label{tab:limitsStop}}
 \centering
 \cmsTableResize{
 \begin{scotch}{|c|cccc|xc|c|lll|}
 Mass & \multicolumn{4}{c|}{Requirements} & \multicolumn{2}{c|}{Yields} & Signal & \multicolumn{3}{c|}{$\sigma$ (pb)}  \\
 (\GeVns{})  & \pt    & \ias    & \invbeta  & M &\multicolumn{1}{c}{SM predicted} & data & eff.    & theo. & exp. & obs.   \\ \hline
 \multicolumn{11}{|c|}{Top squark with the \tkonly analysis} \\ \hline
  200 &     65 & 0.3 & \NA &    0 & 28.700,6.030 & 24 & 0.195 & 6.1$\times 10^{+1}$ & 3.3$\times 10^{-2}$ & 2.5$\times 10^{-2}$ \\
  600 &     65 & 0.3 & \NA &   40 & 28.700,6.030 & 24 & 0.266 & 1.7$\times 10^{-1}$ & 2.4$\times 10^{-2}$ & 1.8$\times 10^{-2}$ \\
 1000 &     65 & 0.3 & \NA &  320 &  0.632,0.136 &  0 & 0.260 & 6.0$\times 10^{-3}$ & 4.7$\times 10^{-3}$ & 4.7$\times 10^{-3}$ \\
 1800 &     65 & 0.3 & \NA &  660 &  0.026,0.006 &  0 & 0.163 & 4.6$\times 10^{-5}$ & 7.4$\times 10^{-3}$ & 7.4$\times 10^{-3}$ \\
 2200 &     65 & 0.3 & \NA &  690 &  0.021,0.005 &  0 & 0.109 & 6.0$\times 10^{-6}$ & 1.1$\times 10^{-2}$ & 1.1$\times 10^{-2}$ \\
\hline
 \multicolumn{11}{|c|}{Top squark charge-suppressed with the \tkonly analysis } \\ \hline
  200 &     65 & 0.3 & \NA &    0 & 28.700,6.030 & 24 & 0.046 & 6.1$\times 10^{+1}$ & 1.4$\times 10^{-1}$ & 1.1$\times 10^{-1}$ \\
  600 &     65 & 0.3 & \NA &   90 & 22.500,4.710 & 16 & 0.169 & 1.7$\times 10^{-1}$ & 3.1$\times 10^{-2}$ & 2.3$\times 10^{-2}$ \\
 1000 &     65 & 0.3 & \NA &  320 &  0.632,0.136 &  0 & 0.195 & 6.0$\times 10^{-3}$ & 7.4$\times 10^{-3}$ & 6.1$\times 10^{-3}$ \\
 1800 &     65 & 0.3 & \NA &  550 &  0.063,0.014 &  0 & 0.124 & 4.6$\times 10^{-5}$ & 9.9$\times 10^{-3}$ & 9.9$\times 10^{-3}$ \\
 2200 &     65 & 0.3 & \NA &  580 &  0.050,0.011 &  0 & 0.087 & 6.0$\times 10^{-6}$ & 1.5$\times 10^{-2}$ & 1.5$\times 10^{-2}$ \\
 \end{scotch}
 }
\end{table*}

\begin{table*}
 \topcaption{
Summary of the search for long-lived tau sleptons: the \pt\,(\GeVns{}),  \ias, \invbeta, and mass thresholds M\,(\GeVns{}) requirements,
the predicted and observed  yields passing these criteria, and the resulting
expected (exp.) and observed (obs.) cross section limits.
The signal efficiencies and theoretical (theo.) cross sections are also listed.
   \label{tab:limitsStau}}
 \centering
 \cmsTableResize{
 \begin{scotch}{|c|cccc|xc|c|lll|}
 Mass & \multicolumn{4}{c|}{Requirements} & \multicolumn{2}{c|}{Yields} & Signal & \multicolumn{3}{c|}{$\sigma$\,(pb)}  \\
 (\GeVns{})& \pt    & \ias    & \invbeta  & M &\multicolumn{1}{c}{SM predicted} & data & eff.    & theo. & exp. & obs.   \\ \hline
 \multicolumn{11}{|c|}{Inclusive tau slepton with the \tktof analysis} \\ \hline
  200 &     65 & 0.175 & 1.25 &   50 &  0.861,0.174 &  0 & 0.290 & 2.8$\times 10^{-1}$ & 6.0$\times 10^{-3}$ & 4.3$\times 10^{-3}$ \\
  308 &     65 & 0.175 & 1.25 &  130 &  0.081,0.016 &  0 & 0.431 & 2.5$\times 10^{-2}$ & 2.9$\times 10^{-3}$ & 2.9$\times 10^{-3}$ \\
  494 &     65 & 0.175 & 1.25 &  260 &  0.008,0.002 &  0 & 0.592 & 1.9$\times 10^{-3}$ & 2.1$\times 10^{-3}$ & 2.1$\times 10^{-3}$ \\
  651 &     65 & 0.175 & 1.25 &  380 &  0.002,0.000 &  0 & 0.662 & 4.1$\times 10^{-4}$ & 1.9$\times 10^{-3}$ & 1.9$\times 10^{-3}$ \\
 1029 &     65 & 0.175 & 1.25 &  610 &  0.000,0.000 &  0 & 0.710 & 2.2$\times 10^{-5}$ & 1.7$\times 10^{-3}$ & 1.7$\times 10^{-3}$ \\
 1599 &     65 & 0.175 & 1.25 &  910 &  0.000,0.000 &  0 & 0.549 & 1.0$\times 10^{-6}$ & 2.3$\times 10^{-3}$ & 2.3$\times 10^{-3}$ \\
\hline
 \multicolumn{11}{|c|}{Direct pair prod. of tau slepton with the \tktof analysis} \\ \hline
  200 &     65 & 0.175 & 1.25 &   40 &  0.924,0.187 &  0 & 0.242 & 8.0$\times 10^{-3}$ & 7.1$\times 10^{-3}$ & 4.9$\times 10^{-3}$ \\
  308 &     65 & 0.175 & 1.25 &  110 &  0.130,0.026 &  0 & 0.315 & 1.5$\times 10^{-3}$ & 3.9$\times 10^{-3}$ & 3.9$\times 10^{-3}$ \\
  494 &     65 & 0.175 & 1.25 &  230 &  0.013,0.003 &  0 & 0.415 & 1.9$\times 10^{-4}$ & 3.0$\times 10^{-3}$ & 3.0$\times 10^{-3}$ \\
  651 &     65 & 0.175 & 1.25 &  330 &  0.003,0.001 &  0 & 0.496 & 4.9$\times 10^{-5}$ & 2.5$\times 10^{-3}$ & 2.5$\times 10^{-3}$ \\
 1029 &     65 & 0.175 & 1.25 &  590 &  0.000,0.000 &  0 & 0.592 & 4.0$\times 10^{-6}$ & 2.0$\times 10^{-3}$ & 2.0$\times 10^{-3}$ \\
 1599 &     65 & 0.175 & 1.25 &  930 &  0.000,0.000 &  0 & 0.504 & 0.0               & 2.5$\times 10^{-3}$ & 2.5$\times 10^{-3}$ \\
 \end{scotch}
 }
\end{table*}

\begin{table*}
 \topcaption{
Summary of the search for long-lived particles from modified Drell--Yan models of various charge: the \pt (\GeVns{}),  \ias, \invbeta, and mass thresholds M (\GeVns{}) requirements,
the predicted and observed  yields passing these criteria, and the resulting
expected (exp.) and observed (obs.) cross section limits.
The signal efficiencies and theoretical (theo.) cross sections are also listed.
   \label{tab:limitsDY}}
 \centering
 \cmsTableResize{
 \begin{scotch}{|c|cccc|xc|c|lll|}
 Mass & \multicolumn{4}{c|}{Requirements} & \multicolumn{2}{c|}{Yields} & Signal & \multicolumn{3}{c|}{$\sigma$ (pb)}  \\
 (\GeVns{})& \pt   & \ias    & \invbeta  & M &\multicolumn{1}{c}{SM predicted} & data & eff.    & theo. & exp. & obs.   \\ \hline
 \multicolumn{11}{|c|}{Modified Drell--Yan $\abs{Q}$ = $1e$ particles with the \tktof analysis} \\ \hline
  200 &     65 & 0.175 & 1.25 &   80 &  0.319,0.065 &  0 & 0.303 & 1.1$\times 10^{-1}$ & 4.2$\times 10^{-3}$ & 4.2$\times 10^{-3}$ \\
  400 &     65 & 0.175 & 1.25 &  210 &  0.018,0.004 &  0 & 0.417 & 7.3$\times 10^{-3}$ & 3.1$\times 10^{-3}$ & 3.1$\times 10^{-3}$ \\
  600 &     65 & 0.175 & 1.25 &  350 &  0.002,0.000 &  0 & 0.461 & 1.2$\times 10^{-3}$ & 2.8$\times 10^{-3}$ & 2.8$\times 10^{-3}$ \\
  800 &     65 & 0.175 & 1.25 &  480 &  0.001,0.000 &  0 & 0.485 & 2.6$\times 10^{-4}$ & 2.6$\times 10^{-3}$ & 2.6$\times 10^{-3}$ \\
 1000 &     65 & 0.175 & 1.25 &  610 &  0.000,0.000 &  0 & 0.485 & 7.6$\times 10^{-5}$ & 2.7$\times 10^{-3}$ & 2.7$\times 10^{-3}$ \\
 1800 &     65 & 0.175 & 1.25 & 1020 &  0.000,0.000 &  0 & 0.312 & 1.0$\times 10^{-6}$ & 4.1$\times 10^{-3}$ & 4.1$\times 10^{-3}$ \\
 2600 &     65 & 0.175 & 1.25 & 1270 &  0.000,0.000 &  0 & 0.114 & 0.0               & 1.1$\times 10^{-2}$ & 1.1$\times 10^{-2}$ \\
\hline
 \multicolumn{11}{|c|}{Modified Drell--Yan $\abs{Q}$ = $2e$ particles with the \tktof analysis} \\ \hline
  200 &     65 & 0.175 & 1.25 &    0 &  0.930,0.188 &  0 & 0.212 & 3.0$\times 10^{-1}$ & 8.0$\times 10^{-3}$ & 6.1$\times 10^{-3}$ \\
  400 &     65 & 0.175 & 1.25 &   90 &  0.230,0.047 &  0 & 0.409 & 2.3$\times 10^{-2}$ & 3.0$\times 10^{-3}$ & 3.0$\times 10^{-3}$ \\
  600 &     65 & 0.175 & 1.25 &  200 &  0.021,0.004 &  0 & 0.481 & 3.5$\times 10^{-3}$ & 2.7$\times 10^{-3}$ & 2.7$\times 10^{-3}$ \\
  800 &     65 & 0.175 & 1.25 &  300 &  0.004,0.001 &  0 & 0.487 & 8.0$\times 10^{-4}$ & 2.6$\times 10^{-3}$ & 2.6$\times 10^{-3}$ \\
 1000 &     65 & 0.175 & 1.25 &  360 &  0.002,0.000 &  0 & 0.449 & 2.4$\times 10^{-4}$ & 2.8$\times 10^{-3}$ & 2.8$\times 10^{-3}$ \\
 1800 &     65 & 0.175 & 1.25 &  410 &  0.001,0.000 &  0 & 0.182 & 4.0$\times 10^{-6}$ & 6.9$\times 10^{-3}$ & 6.9$\times 10^{-3}$ \\
 2600 &     65 & 0.175 & 1.25 &  480 &  0.001,0.000 &  0 & 0.069 & 0.0               & 1.8$\times 10^{-2}$ & 1.8$\times 10^{-2}$ \\
 \end{scotch}
 }
\end{table*}

\begin{table*}
  \caption{Mass limits obtained at $\sqrt{s}=13$\TeV for various HSCP candidate models compared with
earlier results for $\sqrt{s}=7+8$\TeV~\cite{Chatrchyan:2013oca}.
In the model name, ``CS'' stands for charged suppressed interaction model and ``DY'' for Drell--Yan.
The limits for doubly charged
particles are also compared to the earlier CMS results obtained with the `\multicharge' analysis, which was
specifically designed to search for multiply charged particles.
     \label{tab:MassLimits}}
 \centering
  \begin{scotch}{llcc}
  Model                            &  analysis used               & $\sqrt{s}=7+8$ TeV & $\sqrt{s}=13$ TeV \\ \hline
  \multirow{2}{*}{Gluino $f=0.1$}  & \tkonly                      & $M>1320$\GeV        & $M>1610$\GeV       \\
                                   & \tktof                       & $M>1290$\GeV        & $M>1580$\GeV       \\ \hline
  Gluino $f=0.1$ CS                 & \tkonly                      & $M>1230$\GeV        & $M>1580$\GeV       \\ \hline
  \multirow{2}{*}{Gluino $f=0.5$}  & \tkonly                      & $M>1250$\GeV        & $M>1520$\GeV      \\
                                   & \tktof                       & $M>1220$\GeV        & $M>1490$\GeV      \\           \hline
 Gluino $f=0.5$ CS                 & \tkonly                      & $M>1150$\GeV       &  $M>1540$\GeV      \\ \hline
  \multirow{2}{*}{Top squark}            & \tkonly                      & $M> 930$\GeV       & $M>1040$\GeV        \\
                                   & \tktof                       & $M> 910$\GeV       & $M>990$\GeV        \\ \hline
  Top squark CS                            & \tkonly                      & $M> 810$\GeV       & $M>1000$\GeV        \\ \hline
  \multirow{2}{*}{GMSB tau slepton}       & \tktof                       & $M> 430$\GeV       & $M> 490$\GeV        \\
                                   & \tkonly                      & $M> 389$\GeV       & $M> 480$\GeV        \\ \hline
  \multirow{2}{*}{Pair prod. tau slepton} & \tktof                       & $M> 330$\GeV       & $M> 240$\GeV        \\
                                   & \tkonly                      & $M> 180$\GeV       & \NA        \\ \hline
  \multirow{2}{*}{DY $\abs{Q}=1e$}       & \tkonly                      & $M> 640$\GeV       & $M> 510$\GeV       \\
                                   & \tktof                       & $M> 650$\GeV       & $M> 550$\GeV       \\ \hline
  \multirow{3}{*}{DY $\abs{Q}=2e$}       & \multicharge                 &$M> 720$\GeV       &  \NA     \\
                                   & \tkonly                      & $M> 520$\GeV       & $M> 680$\GeV       \\
                                   & \tktof                       & $M> 520$\GeV       & $M> 660$\GeV       \\
  \end{scotch}
\end{table*}

\clearpage
\begin{acknowledgments}
\tolerance=800
We congratulate our colleagues in the CERN accelerator departments for the excellent performance of the LHC and thank the technical and administrative staffs at CERN and at other CMS institutes for their contributions to the success of the CMS effort. In addition, we gratefully acknowledge the computing centers and personnel of the Worldwide LHC Computing Grid for delivering so effectively the computing infrastructure essential to our analyses. Finally, we acknowledge the enduring support for the construction and operation of the LHC and the CMS detector provided by the following funding agencies: BMWFW and FWF (Austria); FNRS and FWO (Belgium); CNPq, CAPES, FAPERJ, and FAPESP (Brazil); MES (Bulgaria); CERN; CAS, MoST, and NSFC (China); COLCIENCIAS (Colombia); MSES and CSF (Croatia); RPF (Cyprus); SENESCYT (Ecuador); MoER, ERC IUT and ERDF (Estonia); Academy of Finland, MEC, and HIP (Finland); CEA and CNRS/IN2P3 (France); BMBF, DFG, and HGF (Germany); GSRT (Greece); OTKA and NIH (Hungary); DAE and DST (India); IPM (Iran); SFI (Ireland); INFN (Italy); MSIP and NRF (Republic of Korea); LAS (Lithuania); MOE and UM (Malaysia); BUAP, CINVESTAV, CONACYT, LNS, SEP, and UASLP-FAI (Mexico); MBIE (New Zealand); PAEC (Pakistan); MSHE and NSC (Poland); FCT (Portugal); JINR (Dubna); MON, RosAtom, RAS and RFBR (Russia); MESTD (Serbia); SEIDI and CPAN (Spain); Swiss Funding Agencies (Switzerland); MST (Taipei); ThEPCenter, IPST, STAR and NSTDA (Thailand); TUBITAK and TAEK (Turkey); NASU and SFFR (Ukraine); STFC (United Kingdom); DOE and NSF (USA).

\hyphenation{Rachada-pisek}
 Individuals have received support from the Marie-Curie program and the European Research Council and EPLANET (European Union); the Leventis Foundation; the A. P. Sloan Foundation; the Alexander von Humboldt Foundation; the Belgian Federal Science Policy Office; the Fonds pour la Formation \`a la Recherche dans l'Industrie et dans l'Agriculture (FRIA-Belgium); the Agentschap voor Innovatie door Wetenschap en Technologie (IWT-Belgium); the Ministry of Education, Youth and Sports (MEYS) of the Czech Republic; the Council of Science and Industrial Research, India; the HOMING PLUS program of the Foundation for Polish Science, cofinanced from European Union, Regional Development Fund, the Mobility Plus program of the Ministry of Science and Higher Education, the National Science Center (Poland), contracts Harmonia 2014/14/M/ST2/00428, Opus 2013/11/B/ST2/04202, 2014/13/B/ST2/02543 and 2014/15/B/ST2/03998, Sonata-bis 2012/07/E/ST2/01406; the Thalis and Aristeia programs cofinanced by EU-ESF and the Greek NSRF; the National Priorities Research Program by Qatar National Research Fund; the Programa Clar\'in-COFUND del Principado de Asturias; the Rachadapisek Sompot Fund for Postdoctoral Fellowship, Chulalongkorn University and the Chulalongkorn Academic into Its 2nd Century Project Advancement Project (Thailand); and the Welch Foundation, contract C-1845.
\end{acknowledgments}
\bibliography{auto_generated}

\cleardoublepage \appendix\section{The CMS Collaboration \label{app:collab}}\begin{sloppypar}\hyphenpenalty=5000\widowpenalty=500\clubpenalty=5000\textbf{Yerevan Physics Institute,  Yerevan,  Armenia}\\*[0pt]
V.~Khachatryan, A.M.~Sirunyan, A.~Tumasyan
\vskip\cmsinstskip
\textbf{Institut f\"{u}r Hochenergiephysik,  Wien,  Austria}\\*[0pt]
W.~Adam, E.~Asilar, T.~Bergauer, J.~Brandstetter, E.~Brondolin, M.~Dragicevic, J.~Er\"{o}, M.~Flechl, M.~Friedl, R.~Fr\"{u}hwirth\cmsAuthorMark{1}, V.M.~Ghete, C.~Hartl, N.~H\"{o}rmann, J.~Hrubec, M.~Jeitler\cmsAuthorMark{1}, A.~K\"{o}nig, I.~Kr\"{a}tschmer, D.~Liko, T.~Matsushita, I.~Mikulec, D.~Rabady, N.~Rad, B.~Rahbaran, H.~Rohringer, J.~Schieck\cmsAuthorMark{1}, J.~Strauss, W.~Treberer-Treberspurg, W.~Waltenberger, C.-E.~Wulz\cmsAuthorMark{1}
% ...
\vskip\cmsinstskip
% ...
1:~~Also at Vienna University of Technology, Vienna, Austria\\
% ...

\end{sloppypar}
\end{document}